\documentclass[prl,twocolumn,showpacs,preprintnumbers,amsmath,amssymb]{revtex4}

\usepackage{graphicx}
\usepackage{dcolumn}
\usepackage{bm}

\begin{document}

\title{Optical combs with a crystalline whispering gallery mode resonator}

\author{Anatoliy A. Savchenkov}
\author{Andrey B. Matsko}
\author{Vladimir S. Ilchenko}
\author{Iouri Solomatine}
\author{David Seidel}
\author{Lute Maleki}

\affiliation{OEwaves Inc., 1010 East Union Street, Pasadena, California, 91106}
\date{\today}

\begin{abstract}
We report on the experimental demonstration of a tunable monolithic
optical frequency comb generator. The device is based on the
four-wave mixing in a crystalline calcium fluoride whispering
gallery mode resonator. The frequency spacing of the comb is given
by an integer number of the free spectral range of the resonator.
We select the desired number by tuning the pumping laser frequency
with respect to the corresponding resonator mode. We also observe
interacting optical combs and high-frequency hyperparametric
oscillation, depending on the experimental conditions. A potential
application of the comb for generating narrowband frequency
microwave signals is demonstrated.
\end{abstract}

\maketitle

Optical combs \cite{udem02n} have become an important tool in a
variety of applications, ranging from metrology to spectroscopy.
Optical combs are usually realized with modulated light from
continuous-wave lasers \cite{kourogi93jqe,ye97ol}, as well as mode
locked lasers \cite{jones00s}. Recently, optical combs produced by
the interaction between a continuous-wave pump laser of a known
frequency with the modes of a monolithic ultra-high-Q whispering
gallery mode (WGM) resonator were demonstrated \cite{delhaye07n}.
The comb generation is based on the four-wave mixing (FWM) process
and hyper-parametric oscillations occurring in the resonator
\cite{kippenberg04prl,savchenkov04prl}. The comb generator is
physically similar to the additive modulational instability ring
laser predicted and demonstrated in the fiber ring resonators
\cite{haelterman92ol,coen97prl,coen01ol}.

A large intracavity intensity in high finesse WGMs enables
four-photon process transforming two pump photons into two sideband
(signal and idler) photons. The sum of frequencies of the generated
photons is equal to twice the frequency of the pumping light
because of the energy conservation law. Increase of the pumping
power leads to a cascading the process, generating multiple
equidistant signal and idler harmonics (optical comb), and also
results in interaction between the harmonics \cite{delhaye07n}.
These optical combs usually have spacing equal to the free spectral
range of the resonators. In this Letter we demonstrate for the
first time generation of tunable optical combs in CaF$_2$
resonators. We have observed combs with $~25 \times m$~GHz ($m$ is
an integer number) frequency spacing in the same resonator. The
spacing (the number $m$) is changed controllably by selecting the
proper detuning of the carrier frequency of the pump laser with
respect to a selected WGM frequency.

The demodulation of the optical comb by means of a fast photodiode
results in the generation of high frequency microwave signals at
comb repetition frequency. This is a consequence, and indeed an
indication, that the comb lines are coherent. The spectral purity
of the signal increases with increasing Q factor of the WGMs, the
optical power of the generated sidebands, and the spectral width of
the comb. We have demonstrated generation of 25~GHz signals with
less than $40$~Hz linewidth and shown that the measured linewidth
value is limited by our experimental setup.

We use WGM resonators with a dense mode spectrum. As the result,
multiple nonlinear optical phenomena coming from interaction of
various mode families have been observed. For instance, some comb
envelops are modulated, and other combs grow asymmetrically. We
have observed generation of stand-alone narrowband signal and idler
sidebands separated by several THz. We show that some phenomena
demonstrated in the overmoded WGM resonators have direct analogies
in optical fibers, while some of them are unique to compact
resonator systems.

In our experiments light from a pigtailed 1550~nm laser was
sent into a CaF$_2$ WGM resonator using one coupling prism, and
was retrieved out of the resonator using a second coupling prism. The
light escaping the prism was collimated and sent into a single
mode fiber. The maximum coupling efficiency was better than 35\%
and was achieved with overcoupling the resonator. Decoupling the
prism resulted in an increase of the quality factor and a decrease of
the light transmission through the resonator. The resonator had a
conical shape with the rounded and polished rim. It had a 2.55~mm
diameter and 0.5~mm thickness. Proper shaping of the resonator
allowed reducing the mode crossection area to less than a hundred of
square microns. The intrinsic Q-factor has been on the order of
$2.5\times 10^9$. The resonator was packaged into a thermally
stabilized box to compensate for external thermal fluctuations.

To achieve a stable comb generation the laser frequency was locked to
a mode of the resonator using Pound-Drever-Hall technique
\cite{drever83apb}. It is important to note that the level and the
phase of the lock is different for the oscillating and
non-oscillating resonators. Increasing the power of the locked
laser above the threshold of the oscillation always resulted in the
lock instability. This is expected since the symmetry of
the resonance changes at the oscillation threshold
\cite{matsko05pra}. We have manually modified the lock parameters
while increasing the laser power that helped us to keep the laser
locked. Modifying the lock parameters we were able to gradually
change the detuning of the laser frequency from the resonance
frequency that led to the modification of the comb.

Our resonator had multiple modes families of high Q whispering
gallery modes. We found that stimulated Raman scattering (SRS)
process has a lower threshold compared with the FWM oscillation
process in the case of the direct pumping of the modes belonging to
the {\em basic} mode sequence. This is an unexpected result because
SRS process has somewhat smaller threshold compared with the
hyperparametric oscillation in the modes having identical
parameters \cite{savchenkov04prl}. The discrepancy is resolved if
we note that different mode families have different quality factors
given by the field distribution in the mode and position of the
couplers. The setup was arranged in such a way that the basic
sequence of the WGMs had lower Q-factor (higher loading) compared
with the higher order transverse modes. The SRS process starts in
the higher-Q modes even though the modes have larger volume ${\cal
V}$. This happens because the SRS threshold power is inversely
proportional to ${\cal V} Q^2$.

\begin{figure}[htb]
\centerline{\includegraphics[width=8.5cm]{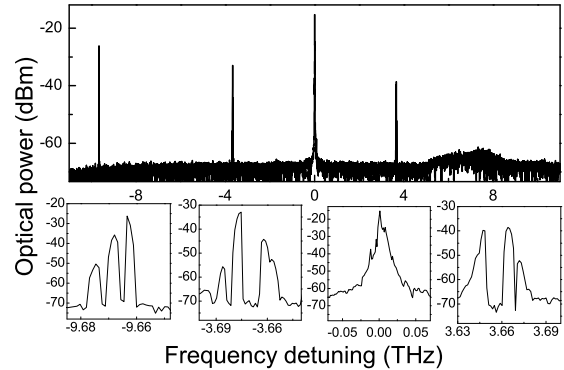}}
\caption{\label{fig1} Frequency spectrum of the SRS
($\sim$9.67 THz from the carrier) and hyperparametric
oscillations observed in the CaF$_2$ resonator pumped to
a mode belonging to the basic mode sequence. The structure
of the lines is shown below the spectrum. The loaded quality factor is $Q=10^{9}$,
the pump power sent to the modes is $8$~mW.}
\end{figure}

Pumping of the basic mode sequence with the light having larger
power leads to hyperparametric oscillations taking place
along with the Raman process (Fig.~\ref{fig1}). Interestingly, the
oscillation occurs at several terahertz detuning from the
pump carrier frequency. Neither hyperparametric oscillation nor
FWM process between the Raman mode and the carrier are observed in
the vicinity of the carrier. Again, this is an unexpected result that
intuitively contradicts the earlier studies
demonstrating generation of the sidebands in the direct vicinity of
the pump frequency (an FSR away from the pump frequency).

The contradiction can be removed if we note that the
hyperparametric process as well as the SRS process start in the
higher Q modes. A study of the signal structure confirms the
conclusion. Indeed, the frequency separation between the modes
participating in the processes is much less than the FSR of the
resonator (see Fig.~\ref{fig1}). The modes are apparently of
the transverse nature. This also explains the absence of the four wave mixing
between the SRS light and the carrier. Nonlinear mixing of the pump
and generated light do not create signals an FSR away from the pump
carrier frequency.

Generation of photon pairs approximately 8~THz apart from the pump
frequency (Fig.~\ref{fig1}) is also intriguing. It seems to be
unclear why such an oscillation frequency is selected by the
system. The question can be answered if we recall the results of
the studies related to phase matching of the FWM process in single
mode \cite{lin81ol} as well as photonic crystal \cite{chen05ol}
fibers in the minimum of the chromatic dispersion region. The dispersion
of the fiber results in unique phase matching conditions for
generation of highly detuned signal and idler if the pumping light
frequency is tuned near zero-dispersion wavelength of the fiber.
The same conditions are valid for our resonators since CaF$_2$ has
its zero dispersion point in the vicinity of 1550~nm
\cite{savchenkov08oe}.

The possibility of generation of photon pairs far away from the
pump makes the WGM resonator-based hyperparametric oscillator
well suited for quantum communication and quantum cryptography
networks. The oscillator avoids large coupling losses occurring
when the photon pairs are launched into communication fibers, in
contrast with the traditional twin-photon sources, based on the
$\chi^{(2)}$ down-conversion process \cite{wang01job}. Moreover,
there is no problem in lossless separation of the narrow band
photons having carrier frequencies several terahertz apart.

To observe generation of optical combs we locked the
frequency of our laser to a transverse WGM. As a result, we
observed hyperparametric oscillation with lower threshold
compared with the SRS process. Even a significant increase of the
optical pump power did not lead to the SRS process because of the
fast growth of the optical combs.

\begin{figure}[htb]
\centerline{\includegraphics[width=7.5cm]{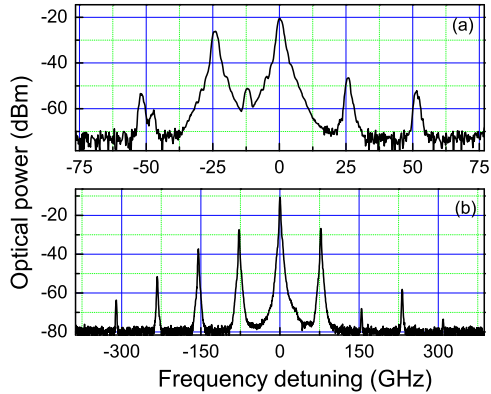}}
\caption{\label{fig2} Hyperparametric oscillation observed in the resonator pumped with 10 mW
of 1550 nm light. Spectra (a) and (b) correspond to different detuning of the pump from
the WGM resonant frequency. One can see the result of the photon summation process in (a), when
the carrier and the first Stokes sideband, separated by 25 GHz, generate photons at 12.5 GHz frequency.
The process is possible because of the high density of the WGMs and is forbidden in the single
mode family resonators.}
\end{figure}

In several cases we observed a significant asymmetry in the
growth of the signal and idler sidebands (see Fig.~(\ref{fig2})).
This asymmetry is not explained with the usual theory of the
hyperparametric oscillation which predicts generation of symmetric
sidebands. We see the explanation in the high modal density in our
resonator. We pump not a single mode, but a nearly degenerate mode
cluster. The transverse mode families have slightly different
geometrical dispersion so the shape of the cluster changes with
the frequency and each mode family results in its own hyperparametric
oscillation. The signal and idler modes of those oscillations are
nearly degenerate so they can interfere. The interference results
in sideband suppression on either side of the carrier. This
results in the "single sideband" oscillations we observe. It is worth
noting, though, that the interfering combs should not be considered
as independent because the generated sidebands have a distinct phase
dependence, as is shown in the later discussion devoted to the
generation of microwave signals by the comb demodulation.

\begin{figure}[htb]
\centerline{\includegraphics[width=7.5cm]{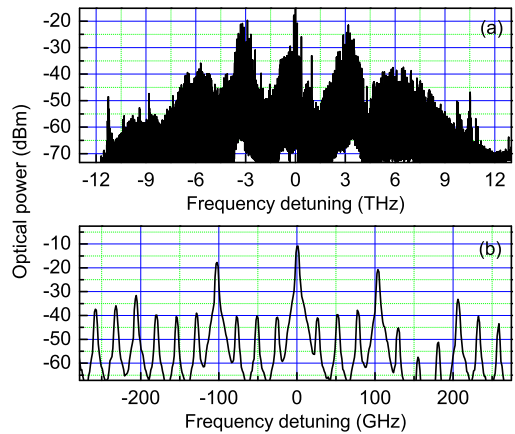}}
\caption{\label{fig3} (a) Optical comb generated by the laser with 50~mW power
(b) represents the central part of (a). The comb has two definite
repetition frequencies equal to one and four FSRs
of the resonator. }
\end{figure}

\begin{figure}[htb]
\centerline{\includegraphics[width=7.5cm]{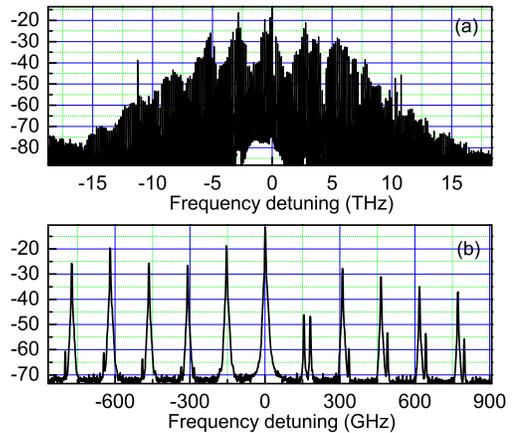}}
\caption{\label{fig4} Modification of the comb shown in Fig.~(\ref{fig3})
when one changes the level and the phase of the lock.}
\end{figure}

The interaction of the signal and idler harmonics becomes even more
pronounced when we increase the pump power to generate optical
combs. We have observed more than 30~THz wide combs in the
resonator (Figs.~(\ref{fig3}) and (\ref{fig4})). The envelops of
the combs are modulated and the reason for the modulation can be
deduced from the Fig.~(\ref{fig4}b). One can see that the comb is
generated over a mode cluster that changes its shape with
frequency.

Another and probably the most important funding is related to the
possibility of controllable tuning of the comb repetition frequency
by changing the frequency of the pump laser. Keeping the
all experimental conditions the same we changed the level and
the phase of the laser lock. This modification of the
experimental conditions resulted in the change of the comb. Examples
are presented in Figs.~(\ref{fig2}-\ref{fig4}).

\begin{figure}[htb]
\centerline{\includegraphics[width=7.5cm]{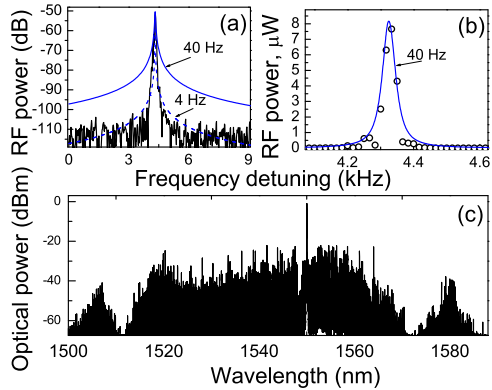}}
\caption{\label{fig5} Microwave signal generated by demodulating the comb with
high frequency photodiode. (a) the signal in the logarithmic scale;
(b) the same signal in the linear scale; (c) optical comb used in the experiment.
Linear fit of the microwave line gives a 40 Hz linewidth, however evaluation of the logarithmic
scale shows that the signal bandwidth is much smaller.}
\end{figure}

To demonstrate the coherent properties of the comb we have sent a
comb having primarily 25~GHz frequency to a fast (40~GHz)
photodiode (optical band 1480-1640~nm) and recorded the microwave
signal. The result of the measurement is shown in
Fig.~(\ref{fig5}). Our microwave spectrum analyzer (Agilent 8564A)
has a 10~Hz video bandwidth, no averaging, and the internal microwave
attenuation is 10~dB (the real microwave noise floor is an order
of magnitude lower). No optical post-filtering of the optical
signal was involved.

It can be seen that the microwave signal is inhomogeneously broadened
to 40~Hz, however the noise floor corresponds to the measurement
bandwidth (is approximately 4 Hz). The broadening comes from the
thermorefractive jitter of the WGM frequency with respect the pump
laser carrier frequency. Our lock is not fast enough
(we use 8~kHz modulation) to compensate
for this jitter. We expect that a better and faster (e. g. 10 MHz)
lock will allow measuring
much narrower bandwidth of the microwave signal. However, even
a 40~Hz linewidth already shows the high coherence of the comb.

It worth to highlight the asymmetric shape of the comb we used in
the microwave experiment Fig.~(\ref{fig5}c). Unlike the nearly
symmetric combs (see Figs.~(\ref{fig3}) and (\ref{fig4})), this
comb is shifted to the blue side of the carrier (c.f.
\cite{delhaye07n}). To produce the comb in Fig.~(\ref{fig5}c) we
locked the laser to one of the modes belonging to the basic mode
sequence. We observed the two mode oscillation process as in
Fig.~(\ref{fig1}) for lower pump power that finally transformed to
the equidistant comb as the power was increased. The SRS process
was suppressed. we expect that the blue shift of the spectrum is
explained by the soliton fission process \cite{husakou01prl}. We
also observed short pulses going out of the oscillator, but our
equipment does not allow resolving pulses shorter than 10~ps. This
issue requires an additional study.

\begin{figure}[htb]
\centerline{\includegraphics[width=7.5cm]{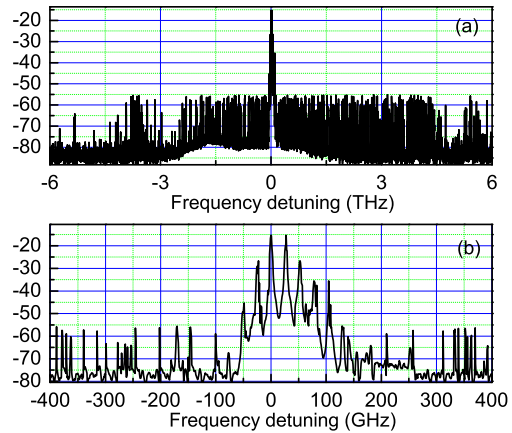}}
\caption{\label{fig6} Harmonics generated by the externally modulated light. We have pumped the
resonator with light 1550 nm light modulated at 25,786~kHz and having 50~mW power. The generated spectrum
is not broader than the spectrum that was produced with a cw pumped resonator and the modes are
not equidistant.}
\end{figure}

Finally, we have sent externally modulated light to the resonator
expecting generation of even broader combs because of the low
dispersion and high nonlinearity of the resonator (c.f. the
continuum generation in microstructured fibers \cite{ranka00ol}).
We have observed the chaotic oscillations instead
(Fig.~(\ref{fig6})). Though the spectrum looked flat and nice, the
generated modes are completely unequidistant. Additional study is
required to explain such a behavior of the system.

To conclude, we studied generation of tunable optical combs in a
WGM crystalline resonator. The combs have spectral widths exceeding
30 THz, as well as outstanding relative coherence of the modes. The
coherence was verified by studying the properties of the microwave
beat note produced by demodulation of the comb on a fast
photodiode. We have shown that the properties of the generated
combs depend significantly on the selection of the optically pumped
mode. We succeeded in locking the pump laser to the WGM to
achieve the stable oscillations in the system.

The authors acknowledge support from the DARPA aPROPOS Program. Andrey
Matsko also acknowledges fruitful discussions with Tobias
Kippenberg.

\end{document}